\documentclass[12pt]{article}
\pdfoutput=1

\usepackage{amsmath,amssymb,amscd}
\usepackage{listings}
\usepackage{caption}
\usepackage{slashed}
\usepackage{color}
\usepackage{ulem}

\usepackage[pdftex]{graphicx}
\usepackage{epstopdf}
\usepackage{epsfig}
\usepackage{listings}
\usepackage{caption}
\usepackage{cite}
\usepackage{amsmath}
\usepackage{breqn}

\newcommand{\beq}{\begin{equation}}
\newcommand{\eeq}{\end{equation}}
\newcommand{\nbea}{\begin{align*}}
\newcommand{\neea}{\end{align*}}
\newcommand{\nbeq}{\begin{equation*}}
\newcommand{\neeq}{\end{equation*}}

\usepackage{array}
\newcolumntype{M}[1]{>{\centering\arraybackslash}m{#1}}
\newcolumntype{N}{@{}m{0pt}@{}}

\numberwithin{equation}{section}

\begin{document}


\pagestyle{empty}

\baselineskip=21pt
\rightline{\footnotesize KCL-PH-TH/2017-21, CERN-TH/2017-084}
\vskip 0.75in

\begin{center}

{\large {\bf Anomaly-Free Dark Matter Models are not so Simple}}

\vskip 0.5in

 {\bf John~Ellis}$^{1,2}$,~
   {\bf Malcolm~Fairbairn}$^{1}$
and {\bf Patrick~Tunney}$^{1}$

\vskip 0.5in

{\small {\it

$^1${Theoretical Particle Physics and Cosmology Group, Physics Department, \\
King's College London, London WC2R 2LS, UK}\\
\vspace{0.25cm}
$^2${Theoretical Physics Department, CERN, CH-1211 Geneva 23, Switzerland}\\
}}

\vskip 0.5in

{\bf Abstract}

\end{center}

\baselineskip=18pt \noindent


{\small
We explore the anomaly-cancellation constraints on simplified dark matter (DM) models with an extra U(1)$^\prime$ gauge
boson $Z'$. We show that, if the Standard Model (SM) fermions are supplemented by a single DM fermion $\chi$ that is a singlet of
the SM gauge group, and the SM quarks have non-zero U(1)$^\prime$ charges, the SM leptons must also
have non-zero U(1)$^\prime$ charges, in which case LHC
searches impose strong constraints on the $Z'$ mass. Moreover, the DM fermion $\chi$ must have a
vector-like U(1)$^\prime$ coupling. If one requires the DM particle to have a purely axial U(1)$^\prime$ coupling, which would be the case if $\chi$ were
a Majorana fermion and would reduce the impact of direct DM searches, the simplest possibility is that it is
accompanied by one other new singlet fermion, but in this case the U(1)$^\prime$ 
charges of the SM leptons still do not vanish.  This is also true in a range of models with multiple
new singlet fermions with identical charges. Searching for a leptophobic model,
we then introduce extra fermions that transform non-trivially under the SM gauge group. We find several such
models if the DM fermion is accompanied by two or more other new fermions with non-identical charges,
which may have interesting experimental signatures. We present benchmark representatives of the various model classes we discuss.}


\vskip 0.75in

\leftline{ {
April 2017}}

\newpage
\pagestyle{plain}

\section{Introduction}

The astrophysical and cosmological necessity for dark matter (DM) (see, for example, \cite{Zwicky,Rubin,Peebles,bullet,planck,dwarf,BertoneHooper}) is one of the strongest motivations for 
particle physics beyond the Standard Model (SM). However, as yet there is no experimental evidence for
any of the proposals for extensions of the SM, such as supersymmetry, that provide well-motivated models
for DM particles \cite{BertoneHooperSilk,EHNOS}. Under these circumstances, a favoured approach is to model dark matter from the bottom up, in other words to avoid
{\it a priori} theoretical assumptions and proceed phenomenologically.  

Initially this programme began by considering higher-dimensional contact interactions \cite{eft1,eft2} where it is straightforward to compare constraints from collider production of dark matter with those from direct detection experiments \cite{moreoneft}.  Such toy models are very useful, but have obvious limitations, since unitarity inevitably breaks down at some scale.  This may, on its own, not be viewed as being problematic in this entirely phenomenological approach, however often the features required to protect unitarity (for example new mediators) themselves lead to interesting phenomenology which is lost in the contact interaction setting \cite{shoemaker,fox,busoni,mccabe}.  The introduction of simplified dark matter models (SDMMs) with the minimal combination of features that a model of DM should have represents an attempt to address this in the simplest way possible~\cite{BM,WhitePapers,tytgat,LHCWG}.
Typically, these SDMMs contain, in addition to the DM particle itself that is often taken to be a fermion, $\chi$, and a bosonic intermediary, $Z'$ (or $\phi$), that generates the interactions between $\chi$ and SM particles and prevents the inherent problems associated with the contact interaction.  

There are then, in general, a number of free parameters associated with the model, for example the masses of the DM and intermediary particles and the seperate couplings of the intermediary to both the DM and SM particles.  One then considers and combines the constraints on these parameters from laboratory experiments at accelerators such as the LHC, direct and indirect astrophysical searches for DM particles, and the allowed range of the cosmological DM density ~\cite{FT,SOMETHINGELSE}. These constraints depend, in particular, whether the DM particle $\chi$ is assumed to be Majorana or Dirac, whether the intermediary has spin zero
or spin one~\footnote{In principle, one could also consider models in which the mediator spin is $\ge 2$, but these have
not yet found much favour.}, in which case it would be associated with an additional U(1)$^\prime$ gauge symmetry,
and whether the mediator couplings are scalar, pseudoscalar, vector or axial vector, all of which have different phenomenologies and constraints \cite{WhitePapers}.

While these simple extensions of the SM are extremely useful for setting up a parameter space which can 
subsequently be  explored, it is well known that many of the simplest models in the SDMM programme are not entirely self-consistent
physically.  For example, models with a massive gauge boson mediator do not respect unitarity to arbitrarily high scales unless set 
within a larger theory where the mass of that boson is explained through an additonal Brout-Englert-Higgs mechanism~\cite{felixkai}.  
Introducing a dark Higgs sector can make such theories more palatable, but the presence of that sector can change the phenomenology of the model.

In this paper we focus on another issue, namely the fact that proposed SDMM extensions to the SM 
with spin-one mediators generally contain anomalies whose cancellation requires additional fermions.
As pointed out in \cite{felixkai}, the masses of the new fermions should be of the same order as the U(1)$^\prime$ boson mass,
offering additional LHC signatures that may already be constrained by the data and should be taken into account in constraining such not-so-SDMMs.

In the case of such a spin-one intermediary particle $Z'$, renormalizability of the SDMM requires that it be free of
anomalous triangle diagrams involving any combination of the SM gauge fields, the U(1)$^\prime$ gauge field
and the graviton~\cite{old}. The requirement of anomaly freedom is understood by constructors of SDMMs~\cite{felixkai,ekstedt}, but in many 
cases its implications have not been pursued fully. One could, of course, take the point of view that any anomalies
in the SDMM could be cancelled by some unspecified ultraviolet completion. However, in this paper we take the
point of view that the SDMM should be self-consistent at the U(1)$^\prime$ scale,
so that one should try to construct anomaly-free SDMMs, and that it is interesting and important to
understand what are the minimal such theories~\footnote{The information gathered in this study may also help
  to guide intuition towards an ultraviolet-complete theory, if one adopts the alternative philosophy.}.

There is a large literature on anomaly-free U(1)$^\prime$ extensions of the SM with various motivations, see for 
example~\cite{barr,erler,appelquist,Carena,batra,morrissey,chiang,langacker,ekstedt,
Duerr,Ismail,Hooper:2014fda,Dobrescu:2014fca,Dudas:2013sia,Wang:2015saa,
Berryman:2016rot,deGouvea:2015pea}.  Among these, the closest in spirit to our
paper are~\cite{Duerr,Ismail}, and we comment later on the relations between their papers and ours.
Typical extensions of the SM with a neutral $Z'$ particle come from GUT theories and couple to leptons as well as quarks \cite{langacker1984}.  
When such a particle acts as the mediator between the SM and a DM fermion, the two strongest constraints come from 
dilepton events at the LHC and direct detection experiments.  

Models in which the $Z'$ boson couples to leptons are very easy to constrain experimentally, since they yield
dilepton events that give clear signals in hadron colliders without the backgrounds that dijets would experience, see for example \cite{tytgat}.
Depending on the model, lower bounds $m_{Z^\prime} \gtrsim 3$~TeV may be imposed by the LHC experiments~\cite{LHCZprime}.
It therefore becomes important to try to suppress the coupling of the mediator particle to the SM leptons for couplings and masses
that give rise to good relic abundance from thermal freeze-out.  This is why one seeks SDMMs containing leptophobic vector mediators that couple only to quarks.

The second very tight constraint comes from the long reach of the latest direct detection experiments - at the time of writing the 
PandaX and LUX experiments have the leading sensitivity to spin-independent dark matter-nucleon scattering,
and have reached cross sections as low as $10^{-46}$~cm$^{2}$ for a DM particle mass of 30 GeV~\cite{PandaX,LUX}.  
This makes it increasingly difficult to arrange couplings and mediator masses that give good relic abundance
and are not ruled out, in the case of a vector mediator interaction that would generate coherent scattering
on all the nucleons in the Xenon nucleus. This coherent scattering is suppressed
by the relative particle velocity if the mediator has an axial coupling to dark matter, and additionally by momentum exchange if it has only axial couplings to quarks~\cite{Kumar:2013iva}~\footnote{ However, we caution that renormalization effects below the U(1)$^\prime$ mass scale
may enhance significantly the scattering of an axially-coupled DM fermion~\cite{dEramo}.}.

The following are the anomaly cancellation conditions involving the U(1)$^\prime$
gauge field that are to be satisfied\footnote{We follow the notation in \cite{ekstedt}: $\mathcal{T}^i$ is a generator of SU(3)$_C$, $T^i$ is a generator of SU(2)$_W$ and $Y$, $Y'$ are hypercharge and U(1)$^\prime$ charge matrices respectively. The U(1) charge matrices are proportional to the identity, but taking the trace will give a factor of two for a doublet relative to a singlet, for example.}, where the trace is over all fermion species
with non-trivial couplings to the corresponding gauge group factors: \\

\begin{itemize}
\item[]{$(\rm a) \quad$ [SU(3)$_C^2$]$\times$[U(1)$^\prime$], which implies Tr[$\{\mathcal{T}^i,\mathcal{T}^j\}Y^\prime$] = 0.}

\item[]{$(\rm b) \quad$ [SU(2)$_W^2$]$\times$[U(1)$^\prime$], which implies Tr[$\{T^i,T^j\}Y^\prime$] = 0.}
\item[]{$(\rm c) \quad$ [U(1)$_Y^2$]$\times$[U(1)$^\prime$], which implies Tr[Y$^2 Y^\prime$] = 0.}
\item[]{$(\rm d) \quad$ [U(1)$_Y$]$\times$[U(1)$^{\prime ^2}$], which implies Tr[$Y Y^{\prime 2}$] =0.}
\item[]{$(\rm e) \quad$ [U(1)$^{\prime 3}$], which implies Tr[$Y^{\prime 3}$] =0.}
\item[]{$(\rm f) \quad$ Gauge-gravity, which implies Tr[$Y$] = Tr[$Y^\prime$] =0.}
\end{itemize}

As we shall see, satisfying these conditions with the DM fermion $\chi$ being the only fermion beyond the SM requires that the U(1)$^\prime$
boson couples to both leptons and quarks, exposing it to sensitive LHC searches, and that the DM fermion has vector-like $Z'$
couplings, placing it within reach of direct searches for DM scattering. A purely axial $\chi - Z'$ coupling is possible only if there are
additional new fermions. The intermediary boson would still have U(1)$^\prime$ couplings to leptons as well as quarks
if there is just one extra singlet fermion, and in a range of models with multiple
new singlet fermions with identical charges.  Continuing the search for a model with vanishing lepton couplings,
we then consider models with extra fermions transforming non-trivially as doublets or triplets of SU(2)$_W$
as well as singlets. We find several classes of such
models if the DM fermion is accompanied by two or more other new fermions with non-identical charges, generalizing a
model presented in~\cite{Duerr}.

\section{New SM Singlet Fermions and Vanishing U(1)$^\prime$ Couplings to Leptons?}
We consider first the possibility that the SDMM contains extra fermions that are singlets under the SM gauge group.
We assume also that the different quark and lepton generations have identical U(1)$^\prime$ charges,
so as to minimize flavour-changing neutral currents. In this case, the anomaly-cancellation conditions
above take the forms~\cite{Carena}: \\
\begin{align}
(\rm a) & \quad 3(2Y^\prime_q - Y^\prime_u - Y^\prime_d) = 0 \, , \label{A1} \\
(\rm b) & \quad 9 Y^\prime_q + 3 Y^\prime_{l}=0 \, , \label{A2} \\
(\rm c) & \quad 2Y^\prime_q - 16 Y^\prime_u - 4 Y^\prime_d + 6 \left( Y^\prime_{l} - 2 Y^\prime_{e} \right) =0 \, , \label{A3} \\
(\rm d) & \quad 6 \left( Y^{\prime 2}_q - 2 Y^{\prime 2}_u + Y^{\prime 2}_d \right) - 6 \left( Y^{\prime 2}_{l} - Y^{\prime 2}_{e} \right) =0 \, , \label{A4} \\
(\rm e) & \quad 9 \left( 2 Y^{\prime 3}_q - Y^{\prime 3}_u - Y^{\prime 3}_d \right) + 3 \left( 2Y^{\prime 3}_{l} - Y^{\prime 3}_{e} \right) + {\rm Tr_{BSM}}(Y^{\prime  3}) =0 \, , \label{A5} \\
(\rm f) & \quad 9 \left( 2 Y^\prime_q - Y^\prime_u - Y^\prime_d\right) + 3 \left( 2Y^\prime_{l} - Y^\prime_{e} \right) + {\rm Tr_{BSM}}(Y^\prime) =0 \, . \label{A6}
\end{align}
where the fermionic U(1)$^\prime$ charges are denoted by $Y^\prime_i$,
$q$ and $l$ label the left-handed quark and lepton doublets, the right-handed fields are labelled $u, d, e$,
and ${\rm Tr_{BSM}}$ denotes a trace over the additional fermions beyond the SM~\footnote{The anomaly-cancellation conditions for the model studied in~\cite{Duerr}
are more complicated, as it has 2 extra U(1) gauge factors, corresponding to baryon and lepton number $B$ and $L$. However, in the limit where one
discards the U(1)$_L$ boson it becomes a leptophobic model with a single U(1)$^\prime$ equivalent to U(1)$_B$, as we discuss later.}.

In the absence of BSM particles, the anomaly cancellation conditions depend only on the $Y^\prime$ charges of the SM fields. The Y-sequential model \cite{appelquist,ekstedt} is a well known example of an anomaly-free $U(1)^\prime$ theory where the $Y'$ charge of each fermion is proportional to the SM $Y$ hypercharge. This solution is trivially guaranteed to exist since the SM is anomaly-free, and so we expect to recover this model in our analysis when ${\rm Tr_{BSM}}(Y^\prime) = {\rm Tr_{BSM}}(Y^{\prime 3}) = 0$. However, this model  has couplings to leptons and hence is subject to the strong LHC dilepton constraints, so first we will see if it is possible to obtain an anomaly-free theory with vanishing couplings to leptons.

In addition to these anomaly cancellation conditions, gauge invariance of the SM Yukawa interactions require, if there is a single Higgs doublet, 
\begin{equation}
Y^\prime_H = Y^\prime_q - Y^\prime_u= Y^\prime_d - Y^\prime_q = Y^\prime_{e} - Y^\prime_{l} \, ,
\label{eq:yukawa}
\end{equation}
where $Y^\prime_H$ is the U(1)$^\prime$ charge of the SM Higgs~\footnote{The conditions (\ref{eq:yukawa}) were not imposed in the models studied in~\cite{Ismail},
which would require multiple Higgs representations in order to accommodate SM fermion masses and quark mixing.}. 
These relations always ensure that the first anomaly condition is satisfied, motivating the
consideration of new fermions that are SU(3) singlets as the simplest possibility. If one does not want to assume a particular mass generation mechanism for the SM fields, we note that equation (\ref{eq:yukawa}) is redundant when equations (\ref{A1})-(\ref{A3}) are solved with exotic fermions transforming trivially with respect to the SM gauge group. As such, our conclusions in this Section and in Section 3 are independent of the Yukawa sector, but we impose (\ref{eq:yukawa}) as independent constraints in Section 4.

We focus first on the second anomaly condition (\ref{A2}) that involves SU(2)$_W$ gauge fields, which we rewrite as:
\begin{equation}
 Y^\prime_{l} = -3 Y^\prime_{q} \, .
 \label{eq:anomlq}
\end{equation}
This equation implies directly that {\it if} $Y^\prime_l = 0$, so as to avoid the strong constraints from dilepton searches at the LHC,
{\it then also} $Y^\prime_{q} = 0$. 
We then consider the second Yukawa condition in (\ref{eq:yukawa}),
namely $Y^\prime_d - Y^\prime_q = Y^\prime_{e} - Y^\prime_{l}$~. {\it If} we now require that $Y^\prime_e = 0$, again
so as to avoid the LHC dilepton constraints, we see that {\it also} $Y^\prime_d =0$ and hence, via the first anomaly condition (\ref{A1}),
{\it also} $Y^\prime_u = 0$. 
We conclude that the boson of a U(1)$^\prime$ model designed to avoid the LHC dilepton constraints would not even
be produced via tree-level quark-antiquark annihilations at the LHC.

Moreover, we note that, if the DM particle $\chi$ is the only new fermion, the fifth and sixth anomaly conditions (\ref{A5}, \ref{A6}) require 
\begin{align}
3 (Y^\prime_u - 4Y^\prime_q)^3 + Y^{\prime 3}_{\chi,L} - Y^{\prime 3}_{\chi,R} &= 0 \, , \label{eq:anom5DM}  \\
3 (Y^\prime_u - 4Y^\prime_q) + Y^\prime_{\chi,L} - Y^\prime_{\chi,R} &= 0 \, , \label{eq:anom6DM} 
\end{align}
to which the only rational solution is $Y^\prime_{\chi,L} = Y^\prime_{\chi,R}$ implying that such a `singleton'
DM particle must have a vector-like U(1)$^\prime$ coupling, but not constraining its magnitude. This solution also implies from (\ref{eq:anom6DM}) that $Y^\prime_u = 4Y^\prime_q$.

To summarize this Section, assuming that the U(1)$^\prime$ charges of the SM fermions are generation-independent,
and that any new fermions that are chirally charged under U(1)$^\prime$ are singlets under the 
SM gauge group, we found that the intermediary U(1)$^\prime$ boson must have leptonic couplings and hence
be subject to LHC searches for dilepton signatures. Moreover, if the DM particle is the only such new singlet
fermion, it must have a vector-like U(1)$^\prime$ coupling.  This would also
be the case if there were other new SM-singlet fermions that are vectorial under U(1)$^\prime$, since they
would not contribute to any of the anomaly equations (\ref{A1}) to (\ref{A6}). This benchmark model has two\footnote{However,
choosing to normalise one of the Y' charges with the freedom to rescale the dark gauge coupling would leave only one free parameter} free coupling parameters,
$Y'_{\chi, L} = Y'_{\chi, R}$ and $Y'_q$, in terms of which the $Z'$ couplings of the other SM fermions and the SM Higgs boson are specified as follows:
\begin{equation}
Y'_l \; = \; -3Y'_q, \quad Y'_e \; = \; -6Y'_q, \quad Y'_d \; = \; -2 Y'_q, \quad Y'_u \; = \; 4 Y'_q,  \quad Y'_H \; = \; -3 Y'_q \, .
\label{Bench1}
\end{equation}
It is possible to scale the overall couplings of the SM and dark sector to the $Z'$ independently, 
although creating a large hierarchy would require accepting the same hierarchy between U(1)$^\prime$ charges.

\section{A DM particle with axial $Z'$ Couplings?}

We now study whether the DM fermion could have an axial $Z'$ coupling if we
allow more new SM-singlet fermions that possess only U(1)$^\prime$ charges, in which case the
constraints from experiments searching directly for DM scattering would be weaker~\cite{dEramo}. We also recall that an axial U(1)$^\prime$ is
the only possibility if the DM particle is a Majorana fermion.

The constraints (\ref{eq:yukawa}) and (\ref{eq:anomlq}) remain valid in this case, so the anomaly conditions (\ref{A1}) to (\ref{A4}) are all satisfied
automatically, and we need only consider the remaining conditions (\ref{A5}, \ref{A6}), which we write in the forms
\begin{align}
3 (Y^\prime_u - 4Y^\prime_q)^3 +\sum_{j} (Y^{\prime 3}_{j,L} - Y^{\prime 3}_{j,R}) &= 0 \, , \label{eq:anom5}  \\
3 (Y^\prime_u - 4Y^\prime_q) + \sum_{j} (Y^\prime_{j,L} - Y^\prime_{j,R}) &= 0 \, , \label{eq:anom6} 
\end{align}
where $Y^\prime_{j,L/R}$ is the U(1)$^\prime$ charge of the left/right-handed component of a new fermion species $j$. 

One obvious solution has $Y^\prime_u =4Y^\prime_q$ and any number of new fermions with $Y^\prime_{j,L} = Y^\prime_{j,R}$. 
In the case of a single new fermion (presumably the DM particle) this is in fact the only solution, as discussed in the previous Section.
It is clear from equations (\ref{eq:anom5}, \ref{eq:anom6}) that if we require a purely axial $Z'$ coupling of the new DM fermion $\chi$, 
we will need at least one other fermion that is charged under U(1)$^\prime$ in order to cancel the DM anomaly contributions.

Therefore, we consider now models that, in addition to a candidate
DM particle $\chi$ with charge $Y^\prime_{\chi,L}=-Y^\prime_{\chi,R}$, contain a single other species $A$ with left- and right-handed
charges $Y^\prime_{A,L}$ and $Y^\prime_{A,R}$ under U(1)$^\prime$ that is also a singlet under the SM group. Solving equations (\ref{eq:anom5}) and (\ref{eq:anom6}) above, we find that
that this last equation can be written as
\begin{equation}
Y^\prime_u = 4 Y^\prime_q - \frac{1}{3}(Y^\prime_{A,L} - Y^\prime_{A,R}) -\frac{2}{3} Y^\prime_{\chi,L} \, .
\end{equation}
Substituting this condition into equation (\ref{eq:anom5}) gives a relatively complicated polynomial equation. Using the arbitrary normalization
$Y^\prime_{\chi,L} = 1$, the solutions we find with U(1)$^\prime$ charges that are the smallest rational numbers are
\begin{eqnarray}
Y^\prime_{A,L} = -1 & , & Y^\prime_{A,R} = 1 \, , \label{axial} \\
Y^\prime_{A,L} =0 & , & Y^\prime_{A,R} = -1 \; \; {\rm or} \; \; Y^\prime_{A,R} = 5/4 \, , \label{axial2} \\
Y^\prime_{A,R} = 0 & , & Y^\prime_{A,L} =-5/4 \; \; {\rm or} \; \; Y^\prime_{A,L} =1 \, ,
\label{firstsolutions}
\end{eqnarray}
where the last pairs of solutions are equivalent, being mirror images. 

In general, there will be mixing between the new neutral fermions $(\chi, A)$ induced by a combination of
`Majorana' mass terms that do not require U(1)$^\prime$ breaking and `Dirac' terms that involve the
intervention of a Higgs vacuum expectation value (vev). As a result, the mass eigenstates will be
orthogonal mixtures of the interaction eigenstates, and the lightest one should be identified as the DM particle.
The pattern of mixing is quite model-dependent, being determined by the assumed pattern of
Majorana-type masses that do not require a Higgs vev as well as the assumed set of Higgs
representations, their vev's and the magnitudes of their couplings. For example, in model (\ref{axial}) above, 
there could be a $2 \times 2$ Majorana-type mass matrix for the $\chi_L$ and $\bar{\chi_R}$, and Dirac terms
due to a Higgs with $Y^'_H = 2$ could extend this to a full rank-4 mass matrix for $\chi_L, \bar{\chi_R}, A_L$ and $\bar{A_R}$.
On the other hand, generating a full rank-4 mass matrix in the first model in (\ref{axial2}) or the second model
in (\ref{firstsolutions}) would also require a Higgs with $Y^'_H = 1$, and obtaining a rank-4 matrix matrix in
the other models in (\ref{axial2}) and (\ref{firstsolutions}) would require also Higgs fields with fractional $Y^'$.

Since the fermion species in the dark sector have different U(1)$^\prime$ charges, they do not respect the
Glashow-Weinberg-Paschos conditions for natural flavour conservation~\cite{GWP}, and the $Z'$ will in general have
off-diagonal interactions with the dark mass eigenstates. The heavier mass eigenstates could therefore decay into the 
DM particle by radiating SM ${\bar f} f$ pairs through a virtual $Z^\prime$~\footnote{This suggests the possibility of an LHC signature that complements the familiar mono-jet/photon/Higgs...
searches, namely one in which an on-shell $Z'$ is produced and decays into the DM particle and a heavier dark
particle whose decay yields a missing energy + dijet final state. In this case there may be no need to require any
initial-state boson radiation.}.
We have identified this DM particle with the $\chi$ interaction eigenstate introduced above, which would indeed be the
lightest mass eigenstate in a suitable degenerate limit of the mass matrix. In this limit it would have a purely
axial U(1)$^\prime$ coupling, and this would also be the case for arbitrary mixing in model (\ref{axial}), where both $\chi$ and $A$
have axial couplings. However, in the cases (\ref{firstsolutions}) the coupling of the lightest mass eigenstate
would not be purely axial if the mixing were non-trivial.

We have searched for all other solutions with rational U(1)$^\prime$ charges
of the form $p/q: |p, q| \in \mathbb{Z}$ and $\le 100$, with the following results
\begin{eqnarray}
Y^\prime_{A,L} = 2 & , & Y^\prime_{A,R} = - \frac{1}{2} \, , \nonumber\\
Y^\prime_{A,L} = - \frac{8}{5} & , & Y^\prime_{A,R} = - \frac{7}{5} \, , \nonumber \\
Y^\prime_{A,L} = \frac{25}{9} & , & Y^\prime_{A,R} =- \frac{29}{9} 
\label{moresolutions}
\end{eqnarray}
and equivalent mirror solutions. However, in all these cases the SM leptons have non-zero U(1)$^\prime$ charges.

We have also explored the possibilities for two or three `generations' of new fermions $X, A$
with `generation'-independent charges. In both cases the first solution in (\ref{firstsolutions}) is again
valid, and in the three-`generation' case there is in addition a solution with $Y^\prime_{A,L} = 0, Y^\prime_{A,R} = 1$
and its mirror. We have not studied the two- and three-`generation' case thoroughly but there are,
in general, fewer solutions within any fixed range of $p$ and $q$ than in the single-`generation' case
(\ref{firstsolutions}, \ref{moresolutions}), and the SM leptons again have non-zero U(1)$^\prime$ charges.

We conclude that, if the DM particle is required to have an axial U(1)$^\prime$ charge so as to minimize
the impacts of DM search experiments, not only will the U(1)$^\prime$ gauge boson again have leptonic
couplings, but also there must be additional fermions with U(1)$^\prime$ charges that could be produced
and detected at the LHC. The simplest solutions have the following U(1)' charges (using the normalization
$Y^\prime_{\chi,L} = - Y^\prime_{\chi,R} = 1$):
\begin{eqnarray}
Y^\prime_{A,L} = -1  &,&  Y^\prime_{A,R} = 1 \, . \label{Bench21} \\
Y^\prime_{A,L} =0 & , & Y^\prime_{A,R} = -1 \quad ({\rm or} \quad Y^\prime_{A,R} = 0 \; , \; Y^\prime_{A,L} =1) \, .
\label{Bench22}
\end{eqnarray}
These two models also have $Y'_q$ as a free parameter, and the remaining SM U(1)$^\prime$ charges are then related (for both models) by the following equation
\begin{eqnarray}
Y'_l & = & -3Y'_q, \quad Y^\prime_u \; = \; 4 Y^\prime_q - \frac{1}{3}(Y^\prime_{A,L} - Y^\prime_{A,R}) -\frac{2}{3} Y^\prime_{\chi,L}, \quad Y'_d \; = \; 2 Y'_q - Y'_u \, , \nonumber \\
Y'_e & = & - 2 Y'_q - Y'_u, \quad Y'_H \; = \; Y'_q - Y'_u .
\label{Bench23}
\end{eqnarray}
DM searches at the LHC are often presented in a way that shows the complementarity between the production of DM and resonant searches for the mediator, for example when comparing missing energy and dijet searches. This presentation is only possible if one is able to treat the dark and visible couplings as independent parameters, which would be possible for (\ref{Bench21}) but not (\ref{Bench22}). This is because the anomaly cancellation in model (\ref{Bench21}) occurs independently in the dark and visible sectors. This allows the dark and SM couplings of the fermions to the $Z'$ to be scaled independently, with the caveat that one would have to be prepared to accept very large or very small charges in order to create a large hierarchy between the dark and visible couplings. On the other hand, anomaly cancellation in model (\ref{Bench22}) relates directly the charges of the dark and visible sectors.

Finally, we recall that only in case (\ref{Bench21}) is the DM particle coupling guaranteed to be purely axial, whatever
the amount of dark fermion mixing.

\section{New Fermions Transforming Non-Trivially under the SM Gauge Group}
In this Section we introduce exotic fermions to cancel the anomalies present in a leptophobic theory. We first build up the minimal field content needed to obtain an anomaly-free solution, before commenting on whether there is still a viable DM candidate $\chi$ present in the theory.

We consider the possibility that there are new fermions transforming under non-trivial representations of the SM gauge group~\footnote{The models
studied in~\cite{Duerr,Ismail} all incorporate fermions that are charged under the SM SU(3)$\times$SU(2).}, 
in which case the question of whether the leptonic U(1)$^\prime$ charges vanish is reopened. In such a case one would also need to
ensure the cancellation of the anomalies involving only SM gauge bosons, which are not listed above. These SM anomalies would vanish if
the fermions are vector-like with respect to the SM gauge group, and then the new fermions would contribute only to the anomalies listed above if they are chiral with respect to the U(1)$^\prime$. This option would open up possibilities for other electroweak signatures,
if they are not too heavy.

In order to analyse this possibility, we first repeat the anomaly conditions (\ref{A1}) to (\ref{A6}) above,
using the Yukawa conditions (\ref{eq:yukawa}) to substitute $Y^\prime_u$ and $Y^\prime_d$, 
and assuming that any new fermions transform in either the trivial or fundamental representations:
\begin{align}
\sum_{f \in SU(3)} (Y^\prime_{f,L}-Y^\prime_{f,R}) &= 0 \, , \label{eq:2anom1} \\
3Y^\prime_l + 9 Y^\prime_q + \sum_{f \in SU(2)} (Y^\prime_{f,L}-Y^\prime_{f,R}) &= 0  \, , \label{eq:2anom2} \\
-6(Y^\prime_l + 3 Y^\prime_q) + \sum_{f} (Y^\prime_{f,L} Y_{f,L}^2 -Y^\prime_{f,R} Y_{f,R}^2) &=0  \, ,  \label{eq:2anom3} \\
12(Y^\prime_e-Y^\prime_l)(Y^\prime_l+3Y^\prime_q) + \sum_{f} (Y^{\prime 2}_{f,L} Y_{f,L} -Y^{\prime 2}_{f,R} Y_{f,R}) &=0  \, , \label{eq:2anom4} \\
-3 \left( Y^{\prime 3}_e -2 Y^{\prime 2}_l (Y^\prime_l-9Y^\prime_q) + 18 Y^{\prime 2}_e Y^\prime_q - 36 Y^\prime_e Y^\prime_l Y^\prime_q \right) + \sum_{f} (Y^{\prime 3}_{f,L} -Y^{\prime 3}_{f,R}) &=0  \, , \label{eq:2anom5} \\
-3Y^\prime_e + 6 Y^\prime_l +  \sum_{f} (Y^\prime_{f,L} -Y^\prime_{f,R}) &=0 \, . \label{eq:2anom6}
\end{align}
The simplest possibility we study is a single new fermion species $A$ that transforms in the fundamental of SU(2) and has
both U(1)$_Y$ and U(1)$^\prime$ charges. In order not to mess up the purely SM anomaly conditions, 
we assume it is vector-like under both SU(2)$_W$ and U(1)$_Y$, so that $Y_{A,L}=Y_{A,R}=Y_A$.
In this case the second and third anomaly cancellation conditions (\ref{eq:2anom2}, \ref{eq:2anom3}) take the form
\begin{align}
3Y^\prime_l + 9Y^\prime_q + Y^\prime_{A,L} - Y^\prime_{A,R} &= 0 \, , \label{eq:21anom2} \\
-6(Y^\prime_l +3Y^\prime_q) + 2 Y_A^2 (Y^\prime_{A,L} - Y^\prime_{A,R}) & = 0 \, . \label{eq:21anom3}
\end{align}
Eliminating $Y^\prime_q$ by substituting (\ref{eq:21anom2}) into (\ref{eq:21anom3}), we find
\begin{equation}
(1+Y_A^2)(Y^\prime_{A,L} - Y^\prime_{A,R}) = 0 \, ,
\end{equation}
which has has only the vector-like solution $Y^\prime_{A,L} = Y^\prime_{A,R}$.  Moreover, in this case $Y^\prime_l + 3Y^\prime_q=0$, so that $Y^\prime_l = 0$ would require $Y^\prime_q = 0$.
Implementing full leptophobia by requiring $Y^\prime_e = 0$ would then require the SM Higgs to have $Y^\prime_H = 0$
and hence also $Y^\prime_u = Y^\prime_d= 0$, again entailing vanishing couplings to quarks.
The same conclusions hold for models with
several new fermion `generations' if their charges are `generation'-independent, or if we had put $A$ in the adjoint representation.

We are therefore led to consider adding another new fermion species $B$ with different SM quantum numbers, 
imposing $Y^\prime_l=Y^\prime_e=0$ in the attempt to find a non-trivial leptophobic solution. If $A$ and $B$ are both doublets (or both triplets) under SU(2),
the only solution is the one with all SM field charges vanishing. Therefore we consider the possibility that $A$
is a doublet under SU(2)$_W$ but $B$ is an SU(2)$_W$ singlet.
In this case the second anomaly (\ref{eq:2anom2}) gives $Y^\prime_q = -\frac{1}{9} (Y^\prime_{A,L} - Y^\prime_{A,R})$ and the sixth anomaly 
(\ref{eq:2anom6}) gives $Y^\prime_{B,R}=Y^\prime_{B,L}+2Y^\prime_{A,L}-2Y^\prime_{A,R}$. Substituting these into the third anomaly (\ref{eq:2anom3}) yields
\begin{equation}
(1+Y_A^2-Y_B^2)(Y^\prime_{A,L}-Y^\prime_{A,R})=0
\end{equation}
We ignore the solution $Y^\prime_{A,L}=Y^\prime_{A,R}$ since it would imply $Y'_q=0$, which would then make all SM charges vanish.  Therefore we must require
\begin{equation}
1+Y_A^2-Y_B^2 = 0;
\end{equation}
to which the only integer solution is $\{0,1\}$. Since we are working in the convention where $Q=T^3+Y/2$,
this solution has half-integer electric charges for both $A$ and $B$, conflicting with the integer charge quantization seen in Nature~\cite{fraction}. We conclude that this solution is not acceptable.

We have also looked for solutions where $A$ is an SU(2)$_W$ triplet. Equation (\ref{eq:2anom2}) is modified, as we are no longer considering fermions solely in the fundamental or trivial representation, becoming
\begin{equation}
9 Y^\prime_q + \sum_{f \in \mathbf{2}} (Y^\prime_{f,L}-Y^\prime_{f,R}) + 4\sum_{f \in \mathbf{3}} (Y^\prime_{f,L}-Y^\prime_{f,R})= 0 \label{eq:2anom2adj}
\end{equation}
where \textbf{2} and \textbf{3} label the fundamental and adjoint representations respectively.
If $B$ is again an SU(2)$_W$ singlet, repeating the same steps as before we find the condition
\begin{equation}
8+3Y_A^2-3Y_B^2 = 0 \, ,
\end{equation}
which has no integer solutions.
Finally, in the case where $A$ is a triplet and $B$ is a doublet we obtain the condition
\begin{equation}
5+3Y_A^2-3Y_B^2 = 0 \, ,
\end{equation}
which also has no integer solutions. Moreover, we have checked that there are still no solutions in these triplet/singlet and
triplet/doublet cases when there are several `generations' of $A$ and $B$ (even with different numbers of each),
as long as the U(1)$^\prime$ charges are `generation'-independent.

We are therefore led to consider models with three or more species of new fermions.
The models studied in~\cite{Duerr,Ismail} all feature six new fermion species. 
However, as already commented, when the U(1)$_L$ is discarded along with its three $\nu_R$ species,
the model studied in~\cite{Duerr} becomes a leptophobic model with a single U(1)$^\prime$ that is equivalent
to U(1)$_B$. In this limit, the new fermions in the model comprise a doublet that is vector-like under SU(2) and has $Y = - 1$, and two singlets with
$Y = - 2, 0$, respectively.\footnote{In our convention of $Q=T_3 + Y/2$, the SM hypercharges are twice those in~\cite{Duerr}.}

We have checked the anomaly-cancellation conditions for other models
containing three new fermion species with different U(1)$_Y$ charges, i.e., 
$A,B,\chi$ in the (SU(2)$_W$, U(1)$_Y$, U(1)$_{Y^\prime}$) representations (2, Y$_A$, Y$^\prime_{A,L/R}$), (1, Y$_B$, Y$^\prime_{B,L/R}$),
and (1, 0, Y$^\prime_{\chi,L/R}$) respectively. In order to obtain a leptophobic solution 
with $Y'_l=Y'_e=0$, the SM Yukawa condition (\ref{eq:yukawa}) imposes $Y'_u=Y'_d=Y'_q$, so we choose $Y'_q$ as the only remaining free SM charge.
Normalizing Y$^\prime_{\chi,L} = 1$, and noting that the SU(3) anomaly condition is satisfied automatically when the Higgs coupling constraint
(\ref{eq:yukawa}) is imposed, the next four anomaly-cancellation conditions yield
\begin{align}
Y'_q &= \frac{1}{9} \left(Y'_{A,R}-Y'_{A,L}\right) \, ,  \label{Yq} \\
Y'_{\chi,R} &= \frac{Y_B^2 \left(2 Y'_{A,L}-2 Y'_{A,R}+1\right)+2 \left(Y_A^2+1\right)
   \left(Y'_{A,R}-Y'_{A,L}\right)}{Y_B^2} \, ,  \label{YCR} \\
Y'_{B,L} &= \frac{\left(Y_A Y_B^3-2 \left(Y_A^2+1\right){}^2\right) Y'_{A,L}+\left(Y_A Y_B^3+2 \left(Y_A^2+1\right){}^2\right) Y'_{A,R}}{2 \left(Y_A^2+1\right) Y_B^2} \, ,  
\label{YBL} \\
Y'_{B,R} &= \frac{\left(Y_A Y_B^3+2 \left(Y_A^2+1\right){}^2\right) Y'_{A,L}+\left(Y_A Y_B^3-2\left(Y_A^2+1\right){}^2\right) Y'_{A,R}}{2 \left(Y_A^2+1\right) Y_B^2}  \, .
\label{YBR}
\end{align}
Using these expressions, the final $U(1)^3$ anomaly condition gives rise to the slightly unwieldy expression:
\begin{dmath}
-\frac{1}{8 \left(Y_A^2+1\right){}^3 Y_B^6} \left[  -16 \left(Y_A^2+1\right){}^3 Y_B^6 \left(\left(Y'\right)_{A,L}^3-\left(Y'\right)_{A,R}^3\right) + \left(\left(Y_A Y_B^3+2 \left(Y_A^2+1\right){}^2\right) Y'_{A,L}+\left(Y_A Y_B^3-2 \left(Y_A^2+1\right){}^2\right) Y'_{A,R}\right){}^3+\left(\left(2 \left(Y_A^2+1\right){}^2-Y_A Y_B^3\right) Y'_{A,L}-\left(Y_AY_B^3+2 \left(Y_A^2+1\right){}^2\right) Y'_{A,R}\right){}^3+8 \left(Y_A^2+1\right){}^3 \left(Y_B^2 \left(2 Y'_{A,L}-2 Y'_{A,R}+1\right)+2 \left(Y_A^2+1\right) \left(Y'_{A,R}-Y'_{A,L}\right)\right){}^3-8 \left(Y_A^2+1\right){}^3 Y_B^6 \right] = 0 \, . \label{eq:horrible}
\end{dmath}
This equation has a symmetry $Y_{A/B} \leftrightarrow -Y_{A/B}$, which facilitates a scan of possible solutions.
We have restricted our search to positive integer values $\le 10$ for $Y_{A/B}$. The other unknowns,
$Y'_{A,L/R}$, are both rational, and we have scanned irreducible rational numbers of the form $\pm p/q$ with $p$ and $q$ 
integers $\le 10$. In order to have integer charge quantisation, and recalling that our convention is $Q = T_3 + Y/2$,
we further require $Y_A$ to be odd (since its a doublet) and $Y_B$ to be even (since its a singlet).

In certain cases (\ref{eq:horrible}) takes a relatively manageable form. One example is for $Y_A = \pm 1$ and $Y_B = \pm 2$,
which is equivalent to the solution discussed in~\cite{Duerr}. In this case, one can either require $Y'_{A,L}=-1$ with 
$Y'_{A,R}$ arbitrary or $Y'_{A,R}=1$ with $Y'_{A,L}$ arbitrary.
The other case is $Y_A = \pm 7$ and $Y_B = \pm 10$, in which case one need only satisfy
\begin{equation}
2 Y'_{A,L}-3 Y'_{A,R}+5 = 0 \quad {\rm or} \quad 3 Y'_{A,L}-2 Y'_{A,R}+5 = 0
\end{equation}
to obtain acceptable solutions.
In addition to these `regular' solutions with a new SU(2)-doublet fermion, we find 26 other `exceptional' solutions that occur in 13 mirror
pairs with $Y'_{A,L} \leftrightarrow - Y'_{A,R}$ that have $Y_{A/B} \le 10$ and $Y'_{A,L/R} = \pm p/q$ with $p, q \le 10$.
The simplest of these is
\begin{equation}
(Y'_{A,L}, Y'_{A,R}, Y_A, Y_B) = (1, \frac{2}{3}, 3, 2) \, , 
\label{Bench31}
\end{equation}
which is accompanied by its mirror solution with $Y'_{A,L} \leftrightarrow - Y'_{A,R}$.

In addition to $Y'_{\chi, L} = 1$ by definition, 
the benchmark solution (\ref{Bench31}) has $Y'_l = Y'_e = 0$ by construction, and hence
\begin{equation}
Y'_q \; = \; Y'_u \; = Y'_d, \quad Y'_H = 0 \, ,
\label{Bench32}
\end{equation}
where $Y'_q$ is fixed by (\ref{Yq}) and the values of $Y'_{\chi, R}, Y'_{B,L/R}$ are fixed by (\ref{YCR}, \ref{YBL}) and (\ref{YBR})
\begin{equation}
Y'_q \; = \; -\frac{1}{27}, \quad Y'_{\chi,R} \; = \; 0, \quad Y'_{B,L} \; = \; -\frac{1}{3}, \quad Y'_{B,R} \; = \; \frac{4}{3} \, .
\end{equation}
We note that this solution admits a small quark charge relative to the DM charge, 
implying good complementarity between dijet and missing energy searches at the LHC.

Finally, we consider the possibilities when $A$ is in the adjoint (triplet) representation of SU(2)$_W$.
In this case, the first four anomaly-cancellation conditions above are modified to
\begin{align}
Y'_q &= -\frac{4}{9} \left(Y'_{A,L}-Y'_{A,R}\right) \, , \label{Yq3} \\
Y'_{\chi,R} &= 3 Y'_{A,L}-3 Y'_{A,R}+Y'_{B,L}-Y'_{B,R}+Y'_{\chi,L} \, , \label{YCR3} \\
Y'_{B,L} &= \frac{\left(3 Y_A Y_B^3+\left(3 Y_A^2+8\right){}^2\right) Y'_{A,R}-\left(\left(3 Y_A^2+8\right){}^2-3 Y_A Y_B^3\right) Y'_{A,L}}{2 \left(3 Y_A^2+8\right) Y_B^2} \, , 
\label{YBL3} \\
Y'_{B,R} &= \frac{\left(3 Y_A Y_B^3+\left(3 Y_A^2+8\right){}^2\right) Y'_{A,L}-\left(\left(3 Y_A^2+8\right){}^2-3 Y_A Y_B^3\right) Y'_{A,R}}{2 \left(3 Y_A^2+8\right) Y_B^2} \, , \label{YBR3}
\end{align}
and the $U(1)^3$ anomaly equation becomes
\begin{dmath}
-\frac{1}{8 \left(3 Y_A^2+8\right){}^3 Y_B^6} \left[ -24 \left(3 Y_A^2+8\right){}^3 Y_B^6
   \left(\left(Y'\right)_{A,L}^3-\left(Y'\right)_{A,R}^3\right)+8 \left(3 Y_A^2+8\right){}^3 \left(Y_B^2
   \left(3 Y'_{A,L}-3 Y'_{A,R}+1\right)-\left(3 Y_A^2+8\right)
   \left(Y'_{A,L}-Y'_{A,R}\right)\right){}^3+\left(\left(3 Y_A Y_B^3+\left(3 Y_A^2+8\right){}^2\right)
   Y'_{A,L}-\left(\left(3 Y_A^2+8\right){}^2-3 Y_A Y_B^3\right) Y'_{A,R}\right){}^3+\left(\left(\left(3
   Y_A^2+8\right){}^2-3 Y_A Y_B^3\right) Y'_{A,L}-\left(3 Y_A Y_B^3+\left(3 Y_A^2+8\right){}^2\right)
   Y'_{A,R}\right){}^3-8 \left(3 Y_A^2+8\right){}^3 Y_B^6 \right] = 0 \, .
\end{dmath}
As before we have the symmetry $Y_{A/B} \rightarrow -Y_{A/B}$. Requiring that $Y_A$ and $Y_B$ are even so as to obtain integer electric charges. 
we identify a set of solutions defined by $Y_A = 0$ and $Y_B= \pm 2$, which satisfy
\begin{equation}
Y'_{A,R} = \frac{1+Y'_{A,L}}{1+3 Y'_{A,L}} \, .
\label{Bench4}
\end{equation}
In addition to $Y'_{\chi, L} = 1$ by definition, $Y_A = 0$ and $Y_B= \pm 2$, and $Y'_{A,L}$ as a free parameter that determines $Y'_{A,R}$
via (\ref{Bench4}), this benchmark solution  again has $Y'_l = Y'_e = 0$ by construction and the conditions
(\ref{Bench32}) are also obeyed, where $Y'_q$ is fixed by (\ref{Yq3}), and the values of $Y'_{\chi, R}, Y'_{B,L/R}$ are fixed 
in this case by (\ref{YCR3}, \ref{YBL3}) and (\ref{YBR3}). Choosing the positive solution $Y_B=2$, this relates the other charges:
\begin{align}
Y'_q &= \frac{4-12Y^{\prime 2}_{A,L}}{9+27 Y'_{A,L}} \\
Y'_{\chi,R} &= \frac{3 Y'_{A,L} (1 + Y'_{A,L})}{1 + 3 Y'_{A,L}} \\
Y'_{B,L} &= \frac{1- 3 Y^{\prime 2}_{A,L}}{1+ 3 Y'_{A,L}} \\
Y'_{B,R} &= - Y'_{B,L}
\end{align}
Picking a specific benchmark with $Y'_{A,L} =1$, for example:
\begin{equation}
Y'_{A,R} = \frac{1}{2}, \quad Y'_{q} = - \frac{2}{9}, \quad Y'_{\chi,R} = \frac{3}{2}, \quad Y'_{B,L} = - \frac{1}{2}, \quad Y'_{B,R} = \frac{1}{2}
\end{equation}
As in the fundamental case, there are also `exceptional' solutions not falling into the class described above.
We find 28 such solutions with $Y_{A/B} \le 10$ and $Y'_{A,L/R} = \pm p/q$ with $p, q \le 10$, occurring in 14 mirror
pairs with $Y'_{A,L} \leftrightarrow - Y'_{A,R}$. The simplest of these is
\begin{equation}
(Y'_{A,L}, Y'_{A,R}, Y_A, Y_B) = (-1, - \frac{3}{2}, 2, 2) \, , 
\end{equation}
which is accompanied by its mirror solution.

Examining the gauge eigenstates, we find no solutions with an axial DM particle $Y'_{\chi,L} = - Y'_{\chi,R}$ in this Section. Therefore, ignoring the possible effects of mixing, we expect strong direct detection bounds to be relevant. However, based on our results in Section~3, we expect that adding two SM-singlet dark fermions would allow an anomaly-free theory to exist in which one of the dark sector particles has an axial coupling.

As in the two-dark-fermion case studied in Section~3, the interaction eigenstates $(A, B, \chi)$ in the models
studied in this Section will in general mix via a combination of `Majorana' and `Dirac' entries in the mass matrix, that
are model-dependent. We do not discuss any details here, but note that many of the remarks made in
Section~3 apply here also: the mixing may give the lightest mass eigenstate (the DM particle) an admixture of 
vector-like coupling, which would vanish in the degenerate limit in which it was much lighter than the other mass
eigenstates, and the heavier mass eigenstates would, in general, decay via off-diagonal $Z'$ couplings into lighter
mass eigenstates by emitting SM ${\bar f}f$ pairs.
Finally we note that ,if the $\chi$ state mixes with a neutral component of $A$ or $B$, then a coupling to the SM $Z$
boson would be generated. Such a coupling is very heavily constrained, see, e.g.,~\cite{1609.09079},
putting pressure on the viability of $\chi$ as a DM candidate in such a case.

\section{Summary}

As we have seen in this paper, the cancellation of anomalies is a non-trivial constraint on SDMMs with a spin-one mediator boson $Z^\prime$.  Our analysis has led us to consider three classes of models:

{\it One Exotic Fermion}

If the SM is supplemented by a single new fermion, a DM particle that is a singlet of the SM gauge group, 
the $Z^\prime$ cannot be leptophobic unless it also decouples from quarks. 
A benchmark model in this class is specified at the end of Section~2, see (\ref{Bench1}).  This model contains a single vector-like fermionic DM candidate which does not contribute to any anomalies -- the assigned charges of the standard model fields alone cancel all anomalies. As such, this model is the Y-sequential model~\cite{appelquist,ekstedt} with the addition of a DM candidate. The relative coupling of the $Z'$ to quarks and leptons is fixed and comparable, meaning that LHC dilepton bounds would rule out much of the parameter space.  Moreover, to
the extent that the DM particle has non-vanishing SM couplings, they must be vectorial, meaning
that the cross section for scattering off a nucleus would not be velocity suppressed and would also be coherently enhanced. 
Therefore an SDMM with just a DM fermion and a $Z^\prime$ is very strongly constrained by LHC searches~\cite{LHCZprime} and direct DM scattering experiments~\cite{PandaX,LUX}.

{\it Axial Dark Matter}

If the DM particle is to have a purely axial U(1)$^\prime$ coupling, which would diminish the impact of the DM scattering experiments~\cite{PandaX,LUX},
then it must be accompanied by at least one other new singlet fermion. However, the U(1)$^\prime$ charges of the SM leptons still do not vanish
if there is a single such fermion, or several with identical charges. Thus, the $Z^\prime$ in such a model would still be subject to strong
LHC constraints~\cite{LHCZprime}. A benchmark model in this class is specified at the end of Section~3, see (\ref{Bench21}) and (\ref{Bench23}). 

{\it Leptophobic Models}

We find several anomaly free leptophobic models only if the DM fermion is accompanied by at least two other new fermions with non-identical charges,
at least one of which is a non-singlet under the SM gauge group. One of these models is the model with a baryonic DM particle presented in~\cite{Duerr}.
These models may be subject to constraints from LHC searches for new fermions with non-trivial SM quantum numbers that would need to be considered
in assessing the parameter spaces of such models.
A benchmark model with a new SU(2)$_W$ doublet fermion is specified in (\ref{Bench31}) and (\ref{Bench32}),
and one with a new SU(2)$_W$ triplet fermion is specified in (\ref{Bench4}).

Beyond the specific models presented here, we re-emphasize the general point that proponents of SDMMs should ensure that they implement the anomaly-cancellation
constraints. The bad news is that the resulting models may not be so simple, but the good news is that anomaly cancellation can relate the SM
and DM couplings of the $Z^\prime$ and furthermore the additional fermions may have novel experimental signatures.

\section*{Acknowledgements}

The work of JE and MF was supported partly by the STFC Grant ST/L000326/1. 
MF and PT are funded by the European Research Council under the European UnionÕs Horizon 2020 programme
(ERC Grant Agreement no.648680 DARKHORIZONS).

\end{document}